\documentclass[a4paper,11pt]{article}
\usepackage{jinstpub} 
\usepackage{lineno}

\newcommand{\sonoarrivatoqui}[1]{\textcolor{red}{\\ \%\%\% sono arrivato qui \%\%\% \\\\}}

\title{\boldmath A new diffuse reflector filament for additive manufacturing of 3D printing finely-segmented plastic scintillator}

 \author[a,1]{A. Krech\note{Corresponding author.},}
 \author[c,d,e]{S. Berns,}
 \author[c,d,e]{E. Boillat,}
 \author[a]{A. Boyarintsev,}
 \author[b]{T. Dieminger,}
\author[a]{B. Grynyov,}
\author[c,d,e]{S. Hugon,}
\author[a]{N. Karavaeva,}
\author[b]{U. Kose,}
\author[b]{B. Li,}
\author[a]{S. Minenko,}
\author[f]{A. De Roeck,}
\author[b]{A. Rubbia,}
\author[b]{D. Sgalaberna,}
\author[a]{T. Sibilieva,}
\author[a]{M. Sibilyev,}
\author[b]{T. Weber,}
\author[b,2]{J. W\"uthrich\note{Now at the University of Zurich, Switzerland},}
\author[b]{X. Zhao}

\affiliation[a]{Institute for Scintillation Materials NAS of Ukraine (ISMA), Nauky ave. 60, 61072 Kharkiv, Ukraine}

\affiliation[b]{ETH Zurich, Institute for Particle Physics and Astrophysics, CH-8093 Zurich, Switzerland}

\affiliation[c]{Haute Ecole Spécialisée de Suisse Occidentale (HES-SO), Route de Moutier 14, CH-2800 Delémont, Switzerland}
\affiliation[d]{Haute Ecole d'Ingénierie du canton de Vaud (HEIG-VD), Route de Cheseaux 1, CH-1401 Yverdon-les-Bains, Switzerland}
\affiliation[e]{COMATEC-AddiPole, Technopole de Sainte-Croix, Rue du Progrès 31, CH-1450 Sainte-Croix, Switzerland}

\affiliation[f]{Experimental Physics Department, CERN, Esplanade des Particules 1, 1211 Geneva 23, Switzerland}

\emailAdd{antonkrech@gmail.com}

\abstract{This study presents the development and the characterisation of a novel white, reflective filament suitable for additive manufacturing of finelysegmented plastic scintillators. The filament is based on polycarbonate (PC) and polymethyl methacrylate (PMMA) polymers loaded with titanium dioxide (TiO$_2$) and polytetrafluoroethylene (PTFE) to enhance reflectivity. A range of filament compositions and thicknesses was evaluated through optical reflection and transmittance measurements of reflective layers made with the Fused Deposition Modelling (FDM) technique. A 3D-segmented plastic scintillator prototype was made with fused injection modelling (FIM) and tested with cosmic-ray muons to assess the light yield and the optical crosstalk. The results of this study demonstrate the feasibility of producing compact, modular 3D-printed scintillator detectors with a performance comparable to that of standard plastic scintillator detectors whilst exhibiting lower light crosstalk and higher light yield, owing to the improved optical properties of the reflector material.}


\arxivnumber{2604.26033} 

\begin{document}
\maketitle
\flushbottom

\section{Introduction}
\label{sec:introduction}

Plastic scintillators are widely used in particle detectors due in part to their fast response and ease of fabrication; they cover a broad range of detector applications including time of flight detectors~\cite{Betancourt2017,Korzenev:2021mny}, tonne-to-kilotonne neutrino active detectors~\cite{Amaudruz:2012esa,Aliaga:2013uqz,MINOS:2008hdf}, sampling calorimeters~\cite{Allan:2013ofa} and scintillating optical fibres~\cite{Joram:2015ymp}. Recently, significant advancements in the development of novel 3D granular scintillating detectors for the imagining of electromagnetic and hadronic showers~\cite{calice}, as well as neutrino interactions~\cite{SoLid:2017ema,Sgalaberna:2017khy}, have been made. Being relatively light-weight with a substantial fraction of hydrogen, along with sub-nanosecond response times, also allows for such detectors to be employed for efficient neutron detection with kinetic energy reconstruction by time of flight~\cite{Agarwal:2022kiv,Gwon:2022bix}.

A 2-tonne plastic scintillator detector made of around 2,000,000 optically-isolated plastic scintillator cubes read out by three orthogonal wavelength-shifting (WLS) fibres is currently collecting accelerator neutrino data at the T2K experiment in Japan. Each single cube was independently manufactured and assembly is a time consuming process ~\cite{KIKAWA2025170616}.

Conventionally, plastic scintillators are manufactured using cast polymerisation \cite{book-plastic}, injection moulding \cite{APPEL2002349}, or extrusion \cite{THEVENIN198053,Pla-Dalmau:2000puk} techniques, followed by mechanical subtractive processing to achieve the desired geometrical shape and precision. However, these methods are labour-intensive and limit the complexity of the detector geometries.
Attempts have been made to simplify detector fabrication, including the development of a prototype of plastic scintillator cubes glued together, achieving a tolerance of  approximately 200~$\mu$m~\cite{Boyarintsev:2021uyw}. This method is suitable for two-dimensional layers, but it is not feasible for the production of a single 3D monolithic volume of plastic scintillator cubes. 

Recent developments in additive manufacturing have introduced new opportunities for producing scintillator elements with high spatial segmentation~\cite{Berns:2020ehg,Sibilieva:2022ket}. In particular, Fused Deposition Modelling (FDM) allows for simultaneous printing of scintillating and reflective materials using two extruders, thereby automating the fabrication of finely segmented detectors~\cite{Berns_2022} without degrading the optical performance of a standard polystyrene (PST) based plastic scintillator~\cite{Berns:2020ehg}.

Previous work~\cite{Berns_2022} demonstrated that filaments composed of polystyrene (PST) and polymethyl methacrylate (PMMA) blended with titanium dioxide (TiO$_2$) could serve as effective reflective components for such applications. While PST-based reflectors exhibit high reflectivity, they are incompatible with PST scintillators due to material mixing during simultaneous printing~\cite{Tanya-funct-materials}. Additionally, a filament of inorganic scintillator can be obtained and used for 3D printing by embedding inorganic granules into a polymer matrix\cite{Sibilieva:2022ket}.
The field of additive manufacturing of scintillator materials is rapidly growing; beyond our work, several developments in 3D printing of plastic scintillators are ongoing on FDM \cite{Lynch_2020}, including neutron-sensitive filaments \cite{Barr_2025},and resins 
\cite{3dprinted-scint-first,KIM2023168537,doi:10.1021/acsapm.2c00316,jne4010019,KIM20202910} and for neutron identification
\cite{Stowell_2021,DOLEZAL2023168602,CHANDLER2023103688}.


Nevertheless, challenges remained in achieving the required transparency and dimensional accuracy necessary for large-scale detector applications. To overcome this, we introduced a hybrid 3D printing approach, named Fused Injection Modelling (FIM), which combines the geometric flexibility of FDM with the precision of injection moulding \cite{supercube-weber,Li:2024txz}.  In FIM, a hollow matrix of reflective material is first printed, and then filled with molten scintillator using a heated injector. This imposes stringent thermal requirements on the reflector filament, which must retain its structural integrity at injection temperatures around $230^{\circ}\text{C}$.  
A prototype comprising of a monolithic block of $5 \times 5 \times 5$ optically-separated 1~cm$^3$ plastic scintillator cubes, refered to as ``SuperCube'' and read out by 2 orthogonal wavelength shifting fibres, was 3D printed using FIM and tested with cosmic-ray muons and test beams. 

In this article, we present the formulation, extrusion, optical characterisation, and detector-level validation of several new white reflective filaments based on polycarbonate (PC) and PMMA matrices loaded with TiO$_2$ and polytetrafluoroethylene (PTFE). The goal is to identify optimal material combinations for high-performance 3D-printed scintillator detectors, to improve both the optical properties and, at the same time, achieve a good tolerance with manufacturing without any subtractive process.

\section{Production and characterisation of the reflective filaments}
\label{sec:reflective-filaments}

\subsection{Production of the reflector filaments}
\label{sec:production-experiment}

\begin{figure}[htbp] 
\centering
		\includegraphics[width=0.48\textwidth]{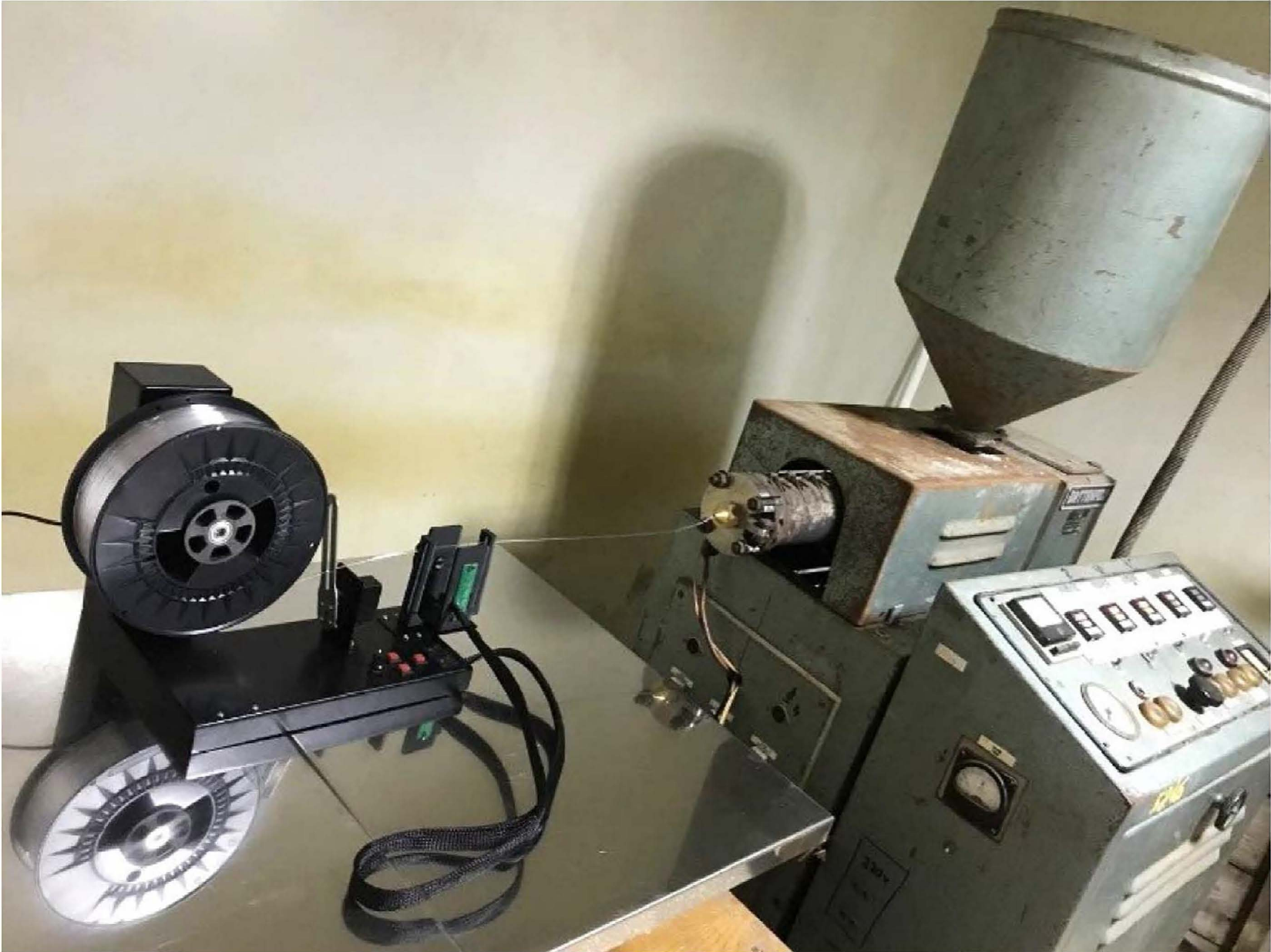}
		\includegraphics[width=0.45\textwidth]{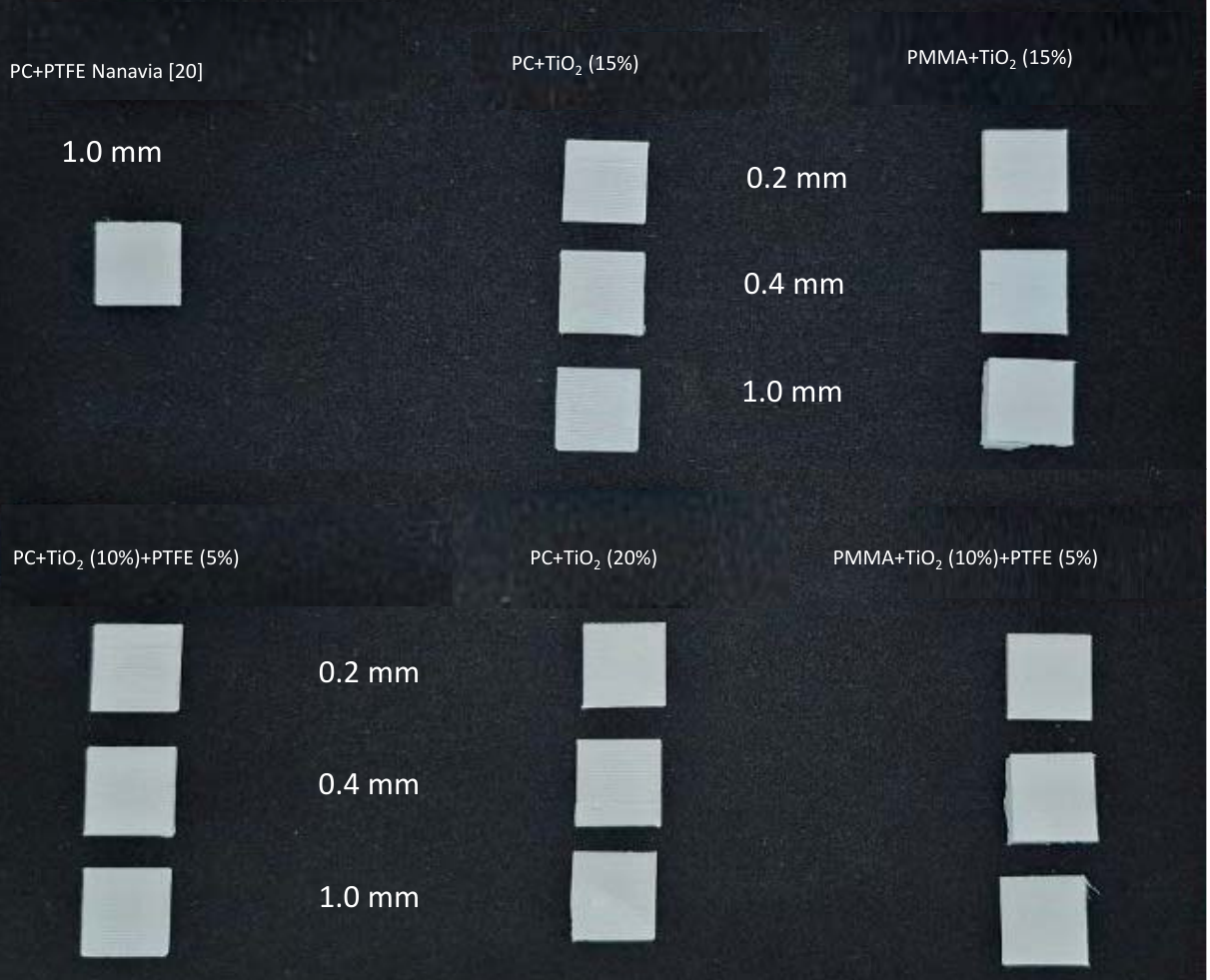}
		\caption{
		Left: the Noztek Filament Winder 1.0 used for large productions of the reflector filaments.
		Right: 3D-printed samples labelled by type of polymer, reflective additive, corresponding concentration, and thicknesses.
		}
		\label{fig:reflectivity}
		\label{fig:photo-3dprint-samples}
\end{figure}

Reflective filaments were fabricated by thermally extruding polymer-additive mixtures using both laboratory and industrial-scale equipment. According to the methodology discussed in \cite{Tanya-funct-materials}, a Noztek ProHT desktop extruder~\cite{extruder-noztek} was used for initial prototyping.  For larger-scale production and consistency checks, an industrial Battenfeld extruder and a Noztek Filament Winder 1.0 \cite{filament-noztek} were used (see Fig.~\ref{fig:reflectivity}).
Polymer granules (PC or PMMA) were first mixed with reflective additive powders, specifically TiO$_2$ and PTFE, in a batch mixer to ensure uniform distribution. Then, the mixture was fed into the extruder, producing a filament with diameter 1.75~$\pm$~0.10~$\text{mm}$, suitable for standard FDM 3D printers.

A Creatbot F430 dual-extruder 3D printer \cite{3dprinter-creatbot-f430} was used to fabricate test samples of reflective material for optical characterisation. Square samples ($20~\times~20~\text{mm}$) with varying thicknesses of 0.2~mm, 0.4~mm, and 1.0~mm were printed. 
The printing temperature was set to $265^{\circ}$ for all samples to ensure consistency across material types and layer heights.
The list of samples and their composition is shown in Fig.~\ref{fig:photo-3dprint-samples}.

The optical properties of the printed samples were measured using a Shimadzu UV-2450 spectrophotometer equipped with an integrating sphere. Measurements of optical reflection and light transmittance were performed over the spectral range of 200–800~$\text{nm}$ \cite{Tanya-funct-materials}. The uncertainty of the instrument on both measurements was 0.5\%.


\subsection{Characterisation of the additive powders}
\label{sec:reflective-powder}

The 
additive powder must maximise reflectivity at wavelengths in the range of the characteristic emission peak of PST-based scintillators, between 400 and 500~$\text{nm}$, while maintaining compatibility with polymer extrusion and printing processes.  

\begin{figure}[htbp]
\centering
		\includegraphics[width=0.65\textwidth]{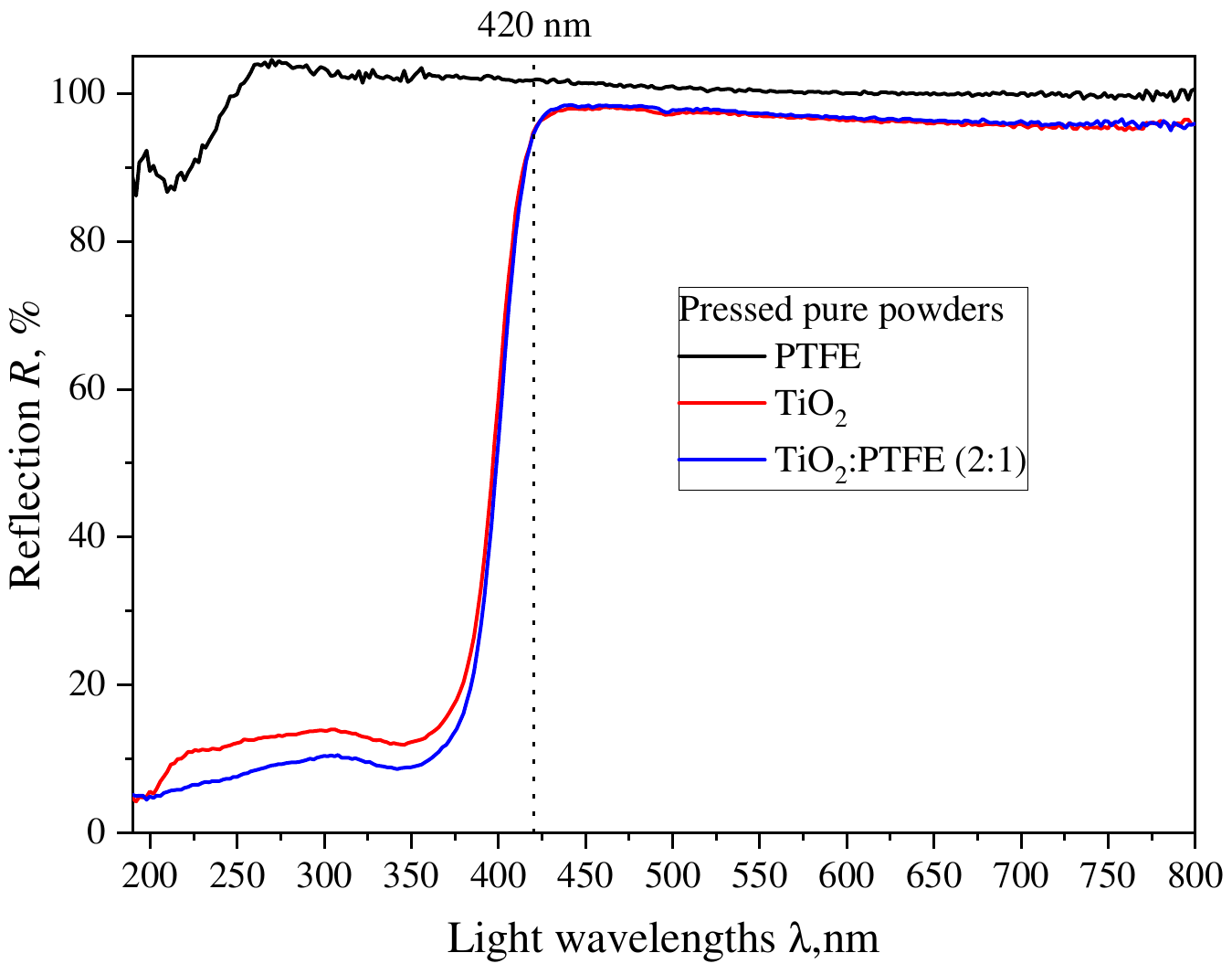}
		\caption{
		Results of the optical reflection (R) measurements performed with pure pressed powders. 
        The reflection value of the reference sample BaSO$_4$ (in the integrating sphere) was taken as 100\%.
		}
		\label{fig:reflection-powders}
\end{figure}

We initially compared the reflectivity of various powdered reflector materials before incorporating them into polymer matrices. 
Pressed powder samples of candidate reflective additives were prepared to compare their pure optical performance independently of the filament polymer matrix.
The results of the reflectivity measurements are summarised in Fig.~\ref{fig:reflection-powders}.
The optical reflection (\textit{R}) was measured with the spectrophotometer Shimadzu-2450. Before each series of measurements, the spectrophotometer is calibrated to minimize the effects of temperature fluctuations and other instrumental variations. A white BaSO$_4$ plate embedded in the integrating sphere is used as the reference sample. During calibration, the signal obtained from the BaSO$_4$ reference is set to 100\% over the entire investigated spectral range. Therefore, the measured reflectance values represent relative reflectance with respect to the BaSO$_4$ reference, which is taken as 100\% at each wavelength. After calibration, an additional control scan of the empty sample compartment (air) is performed over the entire wavelength range. A flat horizontal spectrum at 100\% confirms that the instrument response is correctly normalized across the investigated spectral range. If deviations from this baseline are observed, for example due to fluctuations in the power supply or external light entering the sample compartment, the instrument is recalibrated and the control scan is repeated. Thus, the 100\% reference is defined independently at each wavelength across the entire operating spectral range, rather than at a single specific wavelength.

In the region of scintillation polystyrene luminescence, 
the pure PTFE showed the highest reflectivity, with measured values exceeding 100\%.
Titanium dioxide TiO$_2$ also demonstrated good performance, with a reflectance of 
94.6\% at 420 nm. 
However, combining TiO$_2$ and PTFE in a 2:1 weight ratio did not yield a significant enhancement over the individual components, 
achieving a reflectance of 94.3\% at 420 nm. 
This behaviour may result from the complex interplay between light scattering and absorption in the mixed TiO$_2$-PTFE particulate system. In strongly-scattering particulate media, diffuse reflectance is governed by the effective scattering and absorption properties of the material and can additionally depend on parameters such as particle concentration, layer thickness, and microstructure \cite{Vargas:97}. PTFE is known to exhibit very high diffuse reflectance \cite{Weidner:81}, while the optical response of TiO$_2$ involves both scattering and wavelength-dependent absorption \cite{MURPHY20071326}. Therefore, the reflectance of the TiO$_2$–PTFE mixture reflects the combined effects of scattering and absorption by both components within the particulate system.

\subsection{Performance of the polycarbonate-based filament}
\label{sec:reflective-filament-pc}

\begin{figure}[htbp]
\centering
		\includegraphics[width=0.52\textwidth]{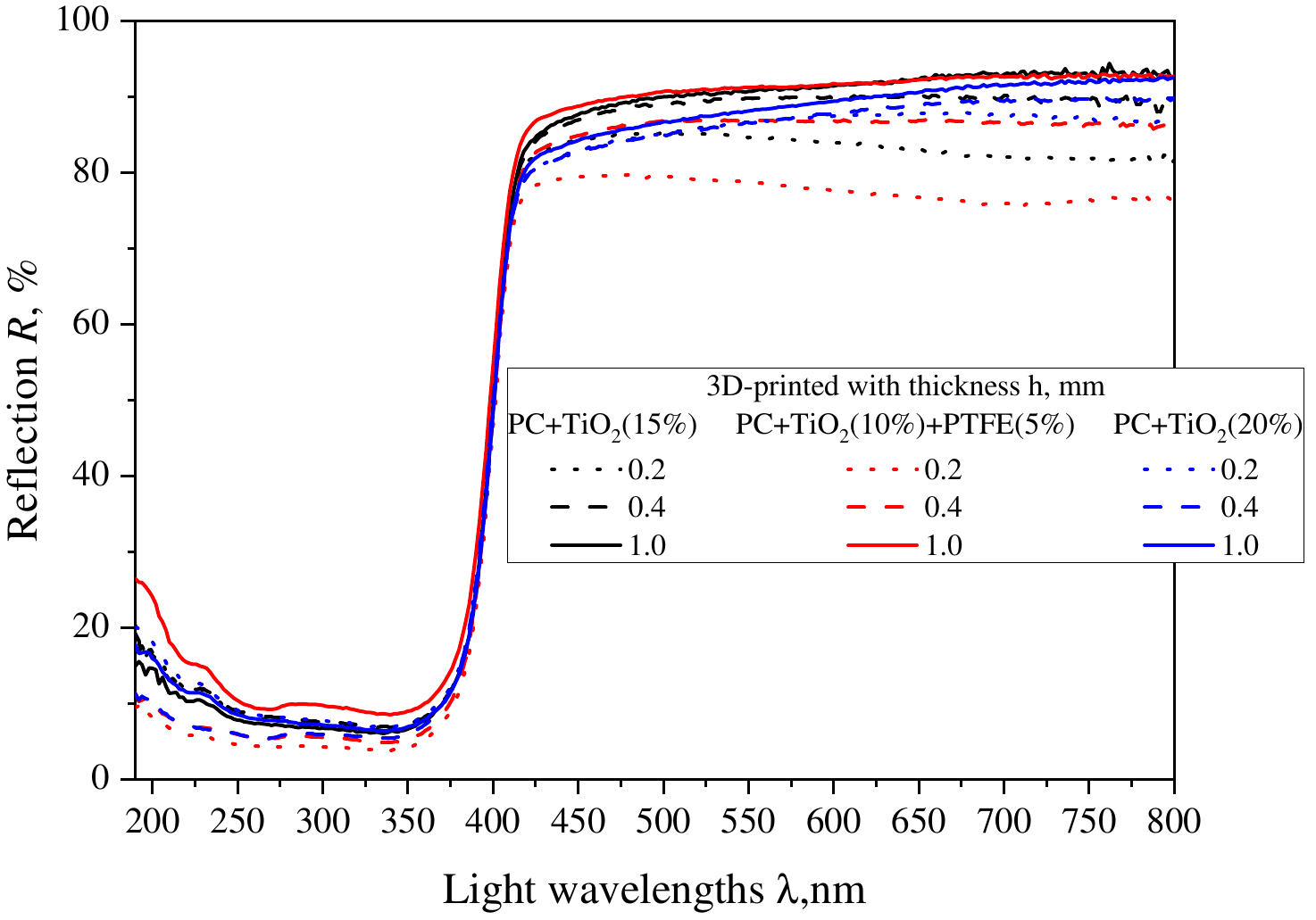}
        \includegraphics[width=0.47\textwidth]{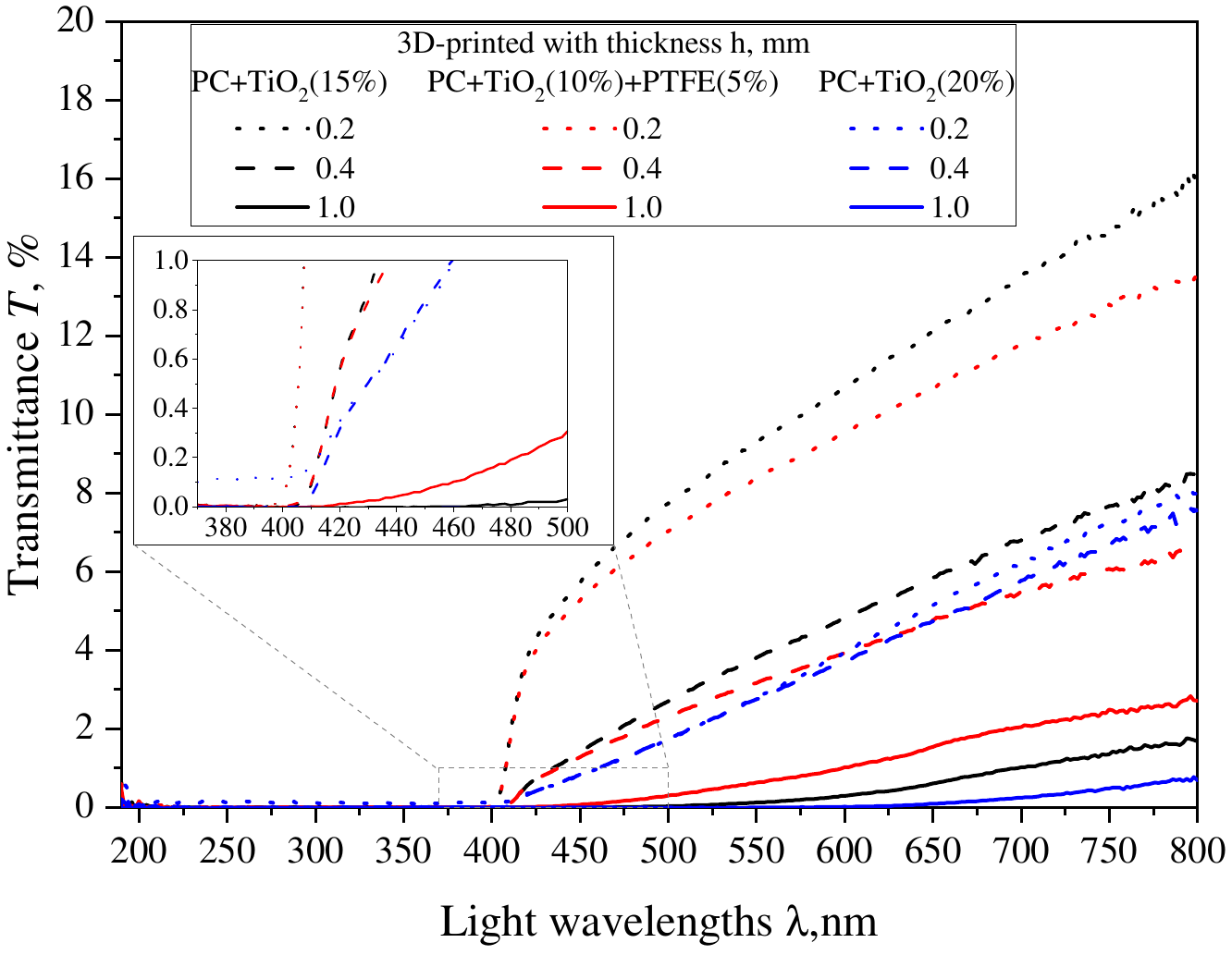}
		\caption{
		Comparison of light reflectivity (left) and transmittance (right) in 3D-printed samples made with PC-based filament with varying concentrations of reflective additives 
        and thicknesses.
		}
		\label{fig:reflection-3dprinted-pc}
        \label{fig:transmittance-3dprinted-pc}
\end{figure}

Following the pure powder reflectivity tests, 
filaments were fabricated using PC as the base polymer, combined with various concentrations of TiO$_2$ and PTFE.
These filaments were used to 3D-print samples of different thicknesses (0.2~mm, 0.4~mm, and 1.0~mm), which were then optically characterised. 
As shown in Fig.~\ref{fig:reflection-3dprinted-pc}, the formulation containing 10\% TiO$_2$ and 5\% PTFE demonstrated high reflectivity in the 410-450~nm wavelength range, with the measured values depending on sample thickness. 
However, increasing TiO$_2$ concentration to 20\% led to decreased performance. This was attributed to the increased brittleness and porosity of the filaments, which negatively impacted the surface smoothness and the internal scattering.

The measurements of the light transmittance 
are shown in Fig.~\ref{fig:transmittance-3dprinted-pc}.
The transmittance reaches the few-percent level already with a thickness of 0.4~mm. Although 0.2~mm-thick samples also provide satisfactory performance, their transmittance is found to be 2–3\% worse than that of the 0.4~mm thick samples. A summary of the results at a wavelength of 420~nm is presented in Tab.~\ref{tab:table-reflection}.
Samples composed of PC with 10\% TiO$_2$ and 5\% PTFE exhibited a lower light transmittance compared to PC + 15\% TiO$_2$ samples. 

\subsection{Performance of the PMMA-based filament}
\label{sec:reflective-filament-pmma}

\begin{figure}[htbp]
\centering
        \includegraphics[width=0.47\textwidth]{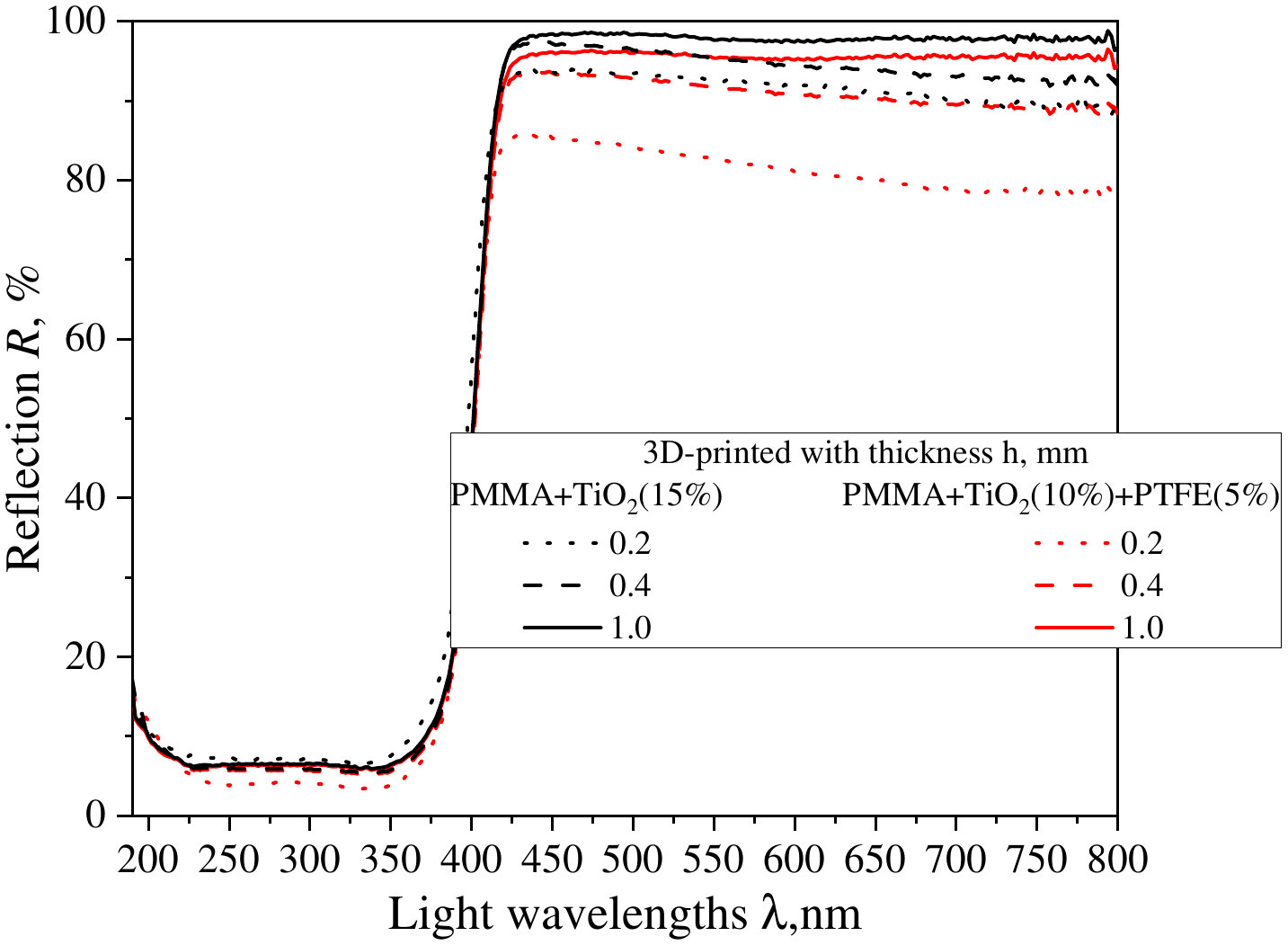}
        \qquad
		\includegraphics[width=0.47\textwidth]{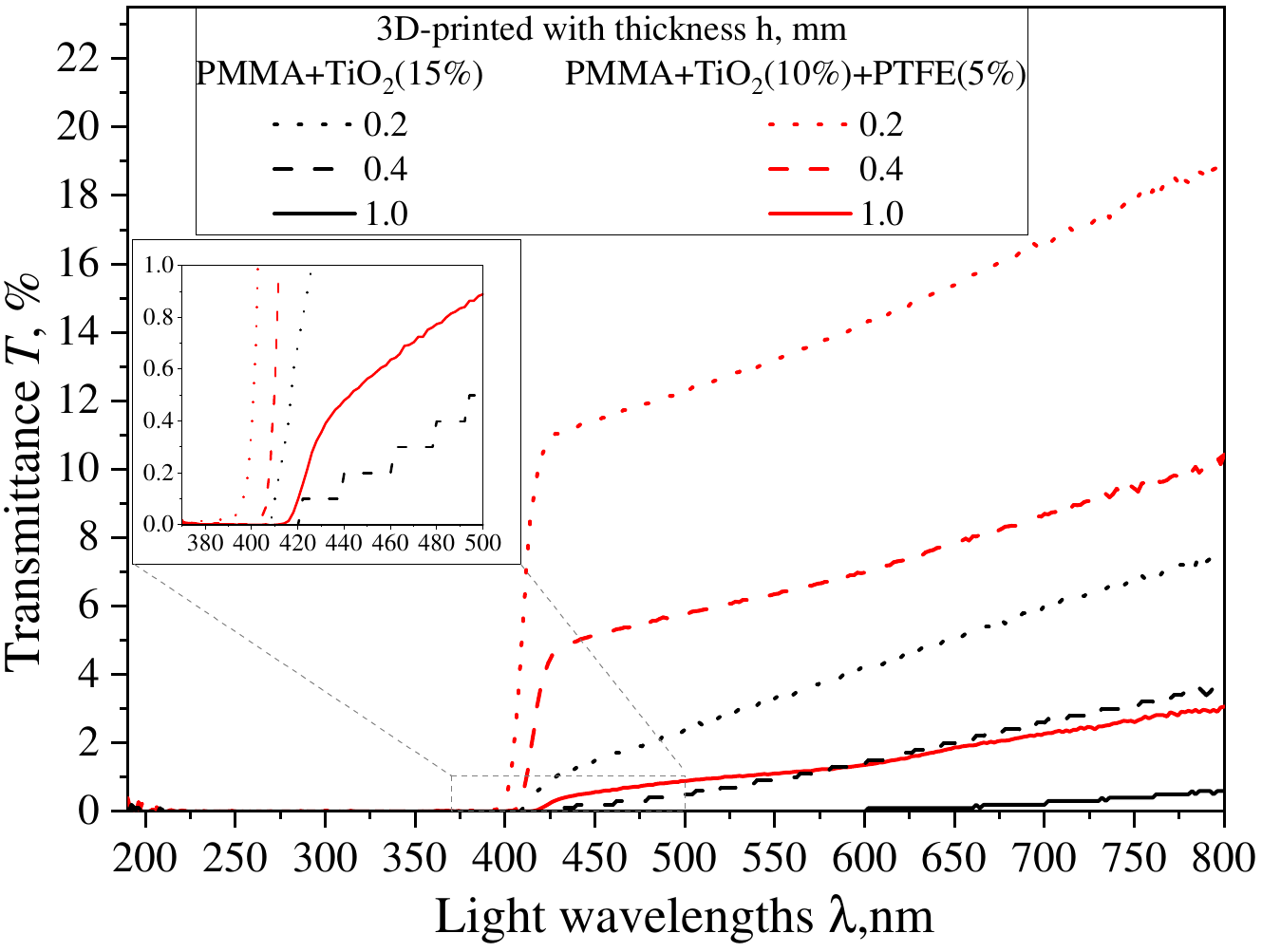}
		\caption{
		Comparison of light reflectivity (left) and transmittance (right) in 3D-printed samples made with PMMA-based filament with varying concentrations of reflective additives and thicknesses.
        }
        \label{fig:comparison-3dprint-pc-vs-pmma}
\end{figure}

To assess the potential of PMMA as an alternative base polymer, the same additive mixture (10\% TiO$_2$ and 5\% PTFE) was introduced into the PMMA filaments. Samples were 3D printed in various thicknesses and characterised under identical conditions as in Sec.~\ref{sec:reflective-filament-pc}. As shown in Fig.~\ref{fig:reflection-3dprinted-pc} and Fig.~\ref{fig:comparison-3dprint-pc-vs-pmma}, PMMA-based samples exhibited both higher reflectivity and higher transmittance compared to PC-based samples. Specifically, at 420~nm, the reflectivity of 0.4~mm thick PMMA samples reached 
90.5\%, 
exceeding that of PC samples. However, this came at the cost of increased light transmittance 
(3.7\% for PMMA vs 0.6\% for PC at 0.4~mm thickness),
which may result in reduced optical isolation between adjacent scintillator voxels in a detector. 

A detailed comparison of reflectivity and transmittance values for both PC and PMMA samples is provided in Tab.~\ref{tab:table-reflection} at 420~nm.
As expected, the lowest light transmittance is observed in 1.0~mm-thick samples due to their greater optical path length. Although, 0.4~mm-thick samples already offer excellent opacity, with transmittance down to 0.3–0.6\%, making them highly suitable for use as reflective material in fine-granularity optically segmented detectors. 
In general, PMMA-based filament shows better reflectivity than that of PC-based ones. 
However, 
this 
is accompanied by a 
higher light transmittance, which would worsen the optical isolation in segmented detector applications. 
For instance, at a thickness of 0.2~mm, PMMA exhibits a transmittance of 10.4\%, compared to only 3.6\% for PC; similarly, at 0.4~mm, the transmittance is 3.7\% for PMMA, whereas PC maintains a much lower value of 0.6\%.
On the other hand, the transmittance is reduced to almost zero for both cases at a sample thickness of 1 mm, while the reflectivity of the PMMA sample remains higher than the PC-based one by 11\%.

\begin{table}[htbp]
\centering
\caption{
Comparison of reflectivity and transmittance for 3D-printed samples of different thicknesses measured at $420$~nm. The performance of the custom filaments is compared with that of the commercial filaments from Rosa3D \cite{rosa3d-filament}, characterised and used to 3D print the SuperCube in \cite{supercube-weber}.
\label{tab:table-reflection}
}
\smallskip
\begin{tabular}{lccc|ccc}
\hline
        \textbf{Sample} & \multicolumn{3}{c|}{\textbf{Reflectivity (\%)}} & \multicolumn{3}{c}{\textbf{Transmittance (\%)}} \\
        \textbf{} & 0.2~mm & 0.4~mm & 1.0~mm & 0.2~mm & 0.4~mm & 1.0~mm \\
        \hline
        PC + TiO$_2$ (10\%) + PTFE (5\%) & 77.6 & 81.2 & 85.5 & 3.6 & 0.6 & 0.0 \\
        PC + TiO$_2$ (15\%) & 81.4 & 83.0 & 83.4 & 3.8 & 0.6 & 0 \\
        PC + TiO$_2$ (20\%) & 79.4 & 79.7 & 80.7 & 0.4 & 0.3 & 0.0 \\
        \hline
        PMMA + TiO$_2$ (10\%) + PTFE (5\%) & 84.7 & 90.5 & 92.1 & 10.4 & 3.7 & 0.1 \\
        PMMA + TiO$_2$ (15\%) & 92.0 & 94.1 & 94.3 & 1.6 & 0.6 & 0.0 \\
        \hline
        PC + PTFE Rosa3D \cite{supercube-weber} & -- & -- & -- & -- & -- & 13.0 \\
        \hline
\end{tabular}
\end{table}

\begin{figure}[htbp]
\centering
		\includegraphics[width=0.45\textwidth]{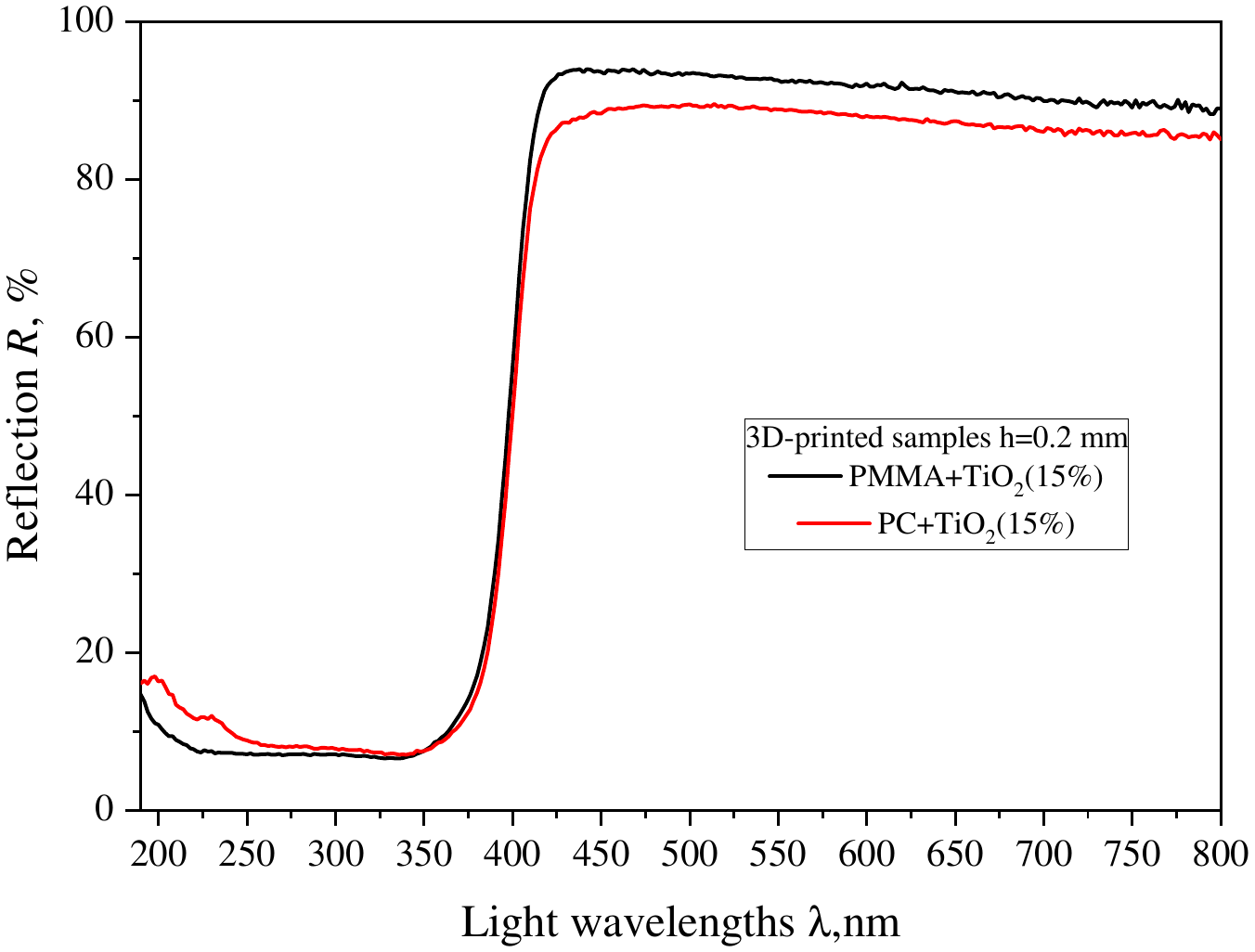}
        \qquad
		\includegraphics[width=0.45\textwidth]{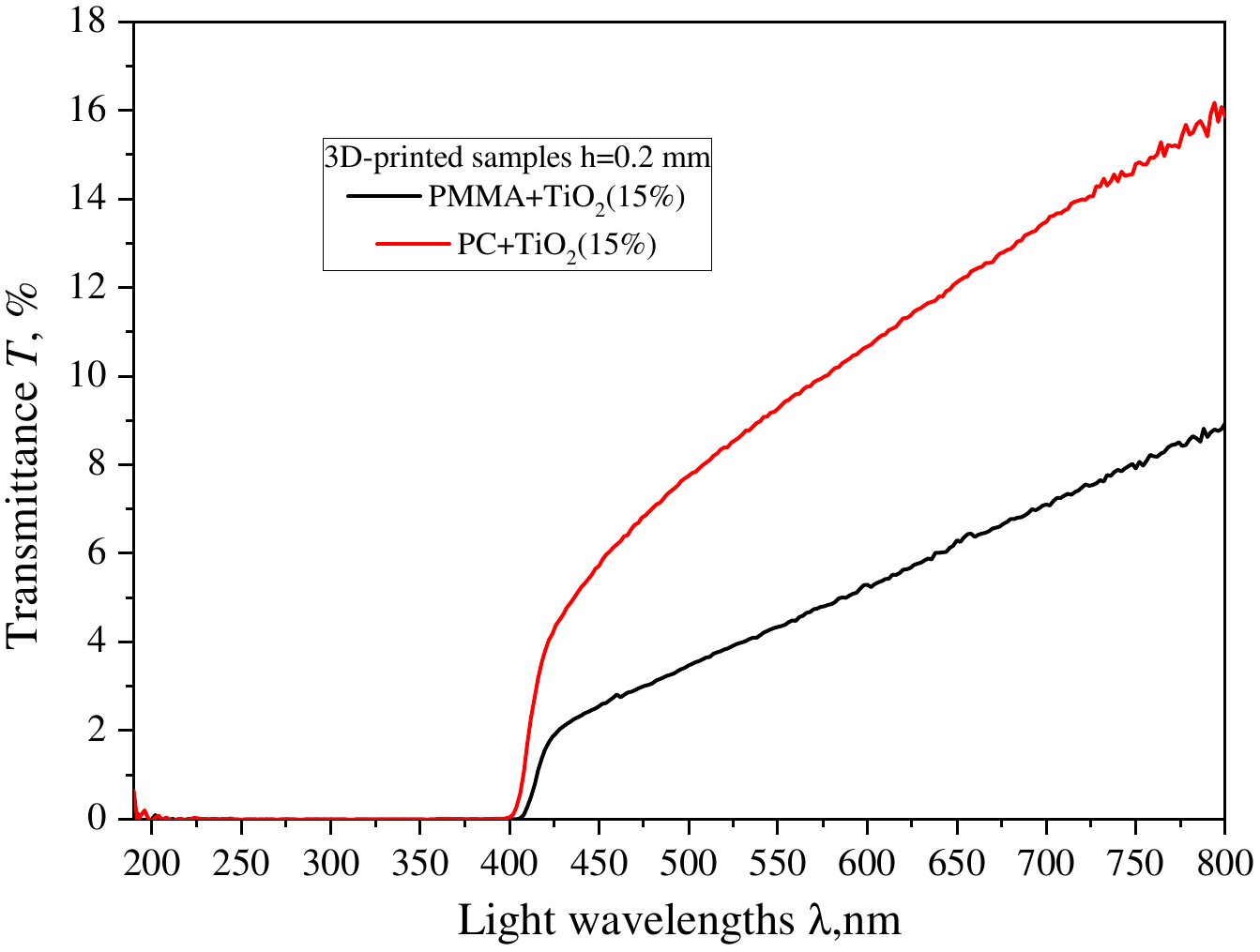}
		\caption{
		Comparison of light reflection (left) and light transmittance (right) in 3D-printed PC and PMMA samples with identical concentrations of reflective additives (15\% TiO$_2$) for a 0.2 mm thickness. 
		}
		\label{fig:comparison-transmittance-reflection-3dprint-pc-vs-pmma-tio2}
\end{figure}

To understand whether the addition of PTFE is necessary in the PMMA formulation, we focus on the filament containing PMMA with 15\% TiO$_2$ only, excluding PTFE. 
The direct comparison of the measured reflection and transmittance values
for PMMA- and PC-based samples without PTFE is presented in Fig.~\ref{fig:comparison-transmittance-reflection-3dprint-pc-vs-pmma-tio2} for a thickness of 0.2~mm.
In this case, PC-based samples exhibited both higher light transmittance and a lower reflectivity than PMMA-based samples.
However, it should also be noted that the softening temperature of PMMA is lower than that of PC. In fact, the typical glass temperature for PMMA ranges between 85~$^{\circ}$C and 105~$^{\circ}$C, while for PC it is approximately 140~$^{\circ}$C.
We can conclude from this that both filament compositions provide a better optical performance than the commercial filament used for the plastic scintillator prototype reported in \cite{supercube-weber}.

\section{Results with the 3D segmented plastic scintillator prototype}
\label{sec:SuperLayer}

The performance of the developed reflector in a 3D segmented plastic scintillator prototype, 3D printed with the FIM technique developed and reported in Ref. \cite{supercube-weber,Li:2024txz} (see Sec.~\ref{sec:introduction} for details), was evaluated.
The prototype, referred to as ``SuperLayer'', 
consists of 4~$\times$~4~$\times$~1  1~cm$^3$ plastic scintillator cubes; the scintillator filament is made of polystyrene 
2\% of \textit{p}-terphenyl (pTP) and 0.05\% of 2,2-p-phenylene-bis(5-phenyloxazole) (POPOP)
by weight.

All six faces of each cube were surrounded by 1~mm thick reflective walls made using the newly developed 3D-printed white filaments (see Sec.~\ref{sec:reflective-filaments}) to ensure an improved optical isolation compared to Ref.~\cite{supercube-weber} and, consequently, to maximise the light collection as well as to reduce the cube-to-cube light crosstalk. 
The scintillation light in each cube was collected by two orthogonal wavelength-shifting (WLS) fibres. 
A total of eight WLS fibres were coupled to Hamamatsu S13360-1325 Multi-Pixel Photon Counters (MPPCs) using black optical connectors.
The SiPM analogue signal was carried via micro-coaxial cables to the CAEN DT5202 \cite{CAEN-DT-5202} front-end board (FEB) used to digitise the analogue signals from the MPPCs.
The pedestal and the gain calibration, the ADC to number of photoelectrons (p.e.) conversion constant, were carried out.
Ref.~\cite{supercube-weber,Li:2024txz} uses the same setup and experimental procedure, and provides more details on the pedestal and gain calibrations. 

\begin{figure}[htbp]
\centering
\includegraphics[width=0.45\textwidth]{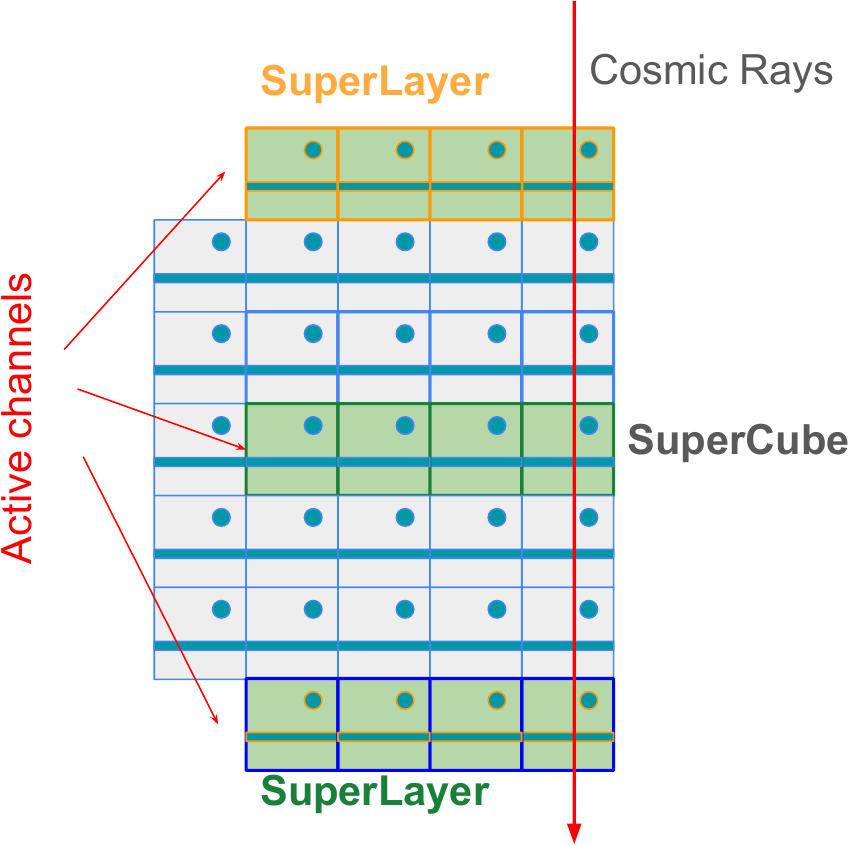}
    \qquad
	\includegraphics[width=0.45\textwidth]{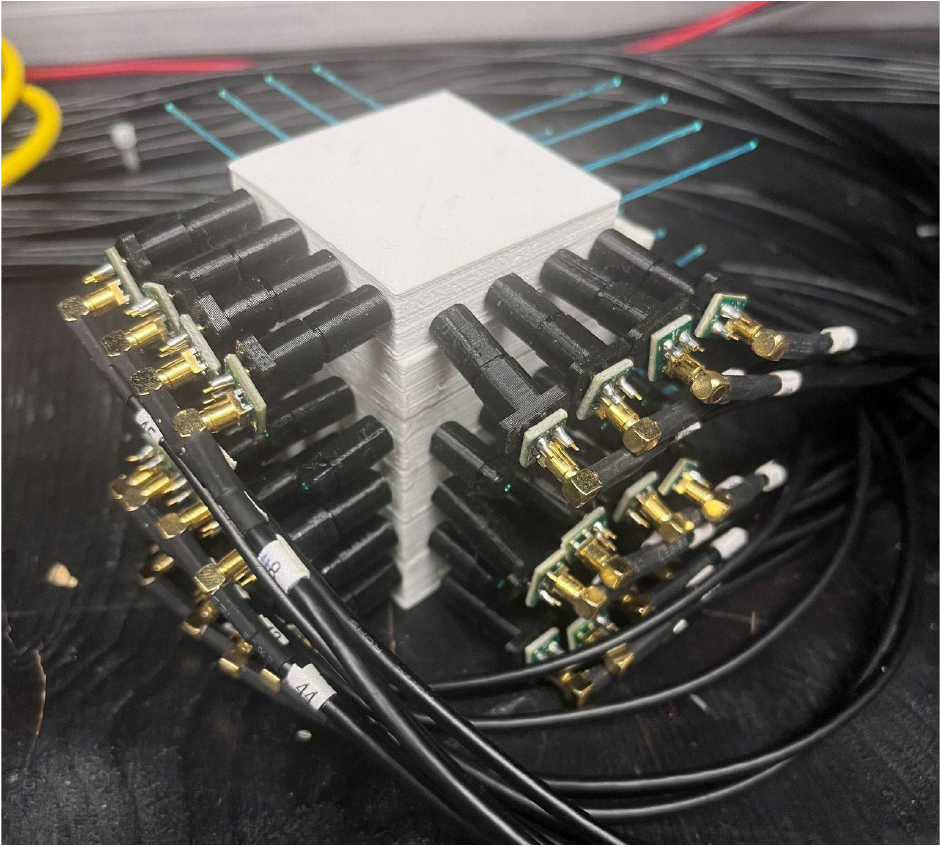}
    \caption{Left: A schematic picture of the measurement setup. The two SuperLayers (green and dark blue lines) are placed right above and below the SuperCube. The green cubes indicate the channels read out during the measurement.
    Right: One SuperLayer stacked on top of the SuperCube with all the WLS fibers inserted and coupled to the SiPMs with black optical connectors. The SiPM analogue signal was carried to the front-end electronics via micro-coaxial cables. 
    } 
	\label{fig:setup}
\end{figure}

Three SuperLayers were 3D printed:
one with the PMMA + 10\% TiO$_2$ + 5\% PTFE filament, another with PC + 10\% TiO$_2$ + 5\% PTFE, and a third with the PMMA + 15 \% TiO$_2$ filament.
It is worth noting that the PMMA-based filaments were easier to 3D print than PC-based filaments, due to the lower melting temperature. For the same reason, the quality of the PMMA-based filament appeared to be visibly better than that of the PC-based filament. This is an important feature in the selection of the optimal filament for a final detector.
The SuperCube prototype (see Sec.~\ref{sec:introduction}) 
was used as a reference and measured together with the SuperLayers. 
The schematic configuration of the measurement, as well as the fully-instrumented prototype setup, is shown in Fig.~\ref{fig:setup}. 
Two SuperLayers were stacked on the top and the bottom of the SuperCube to ensure geometrical symmetry. The 4~$\times$~4 cubes in the middle layer of the SuperCube were measured to provide a baseline reference. 
Vertical-going cosmic-ray particles were selected by requiring signal coincidence in the three cube layers aligned along the same column of cubes, as illustrated. 
In each measurement, around 400 events penetrating the two layers and the SuperCube were selected for analysis,
with the goal of comparing the performance of the old SuperCube, made with 
the same scintillator filament but
a commercial white filament, with that of the new SuperLayers, made with different custom filaments (see full list in Tab.~\ref{tab:table-reflection}).

\begin{figure}[htbp]
\centering 
\includegraphics[width=0.45\textwidth]{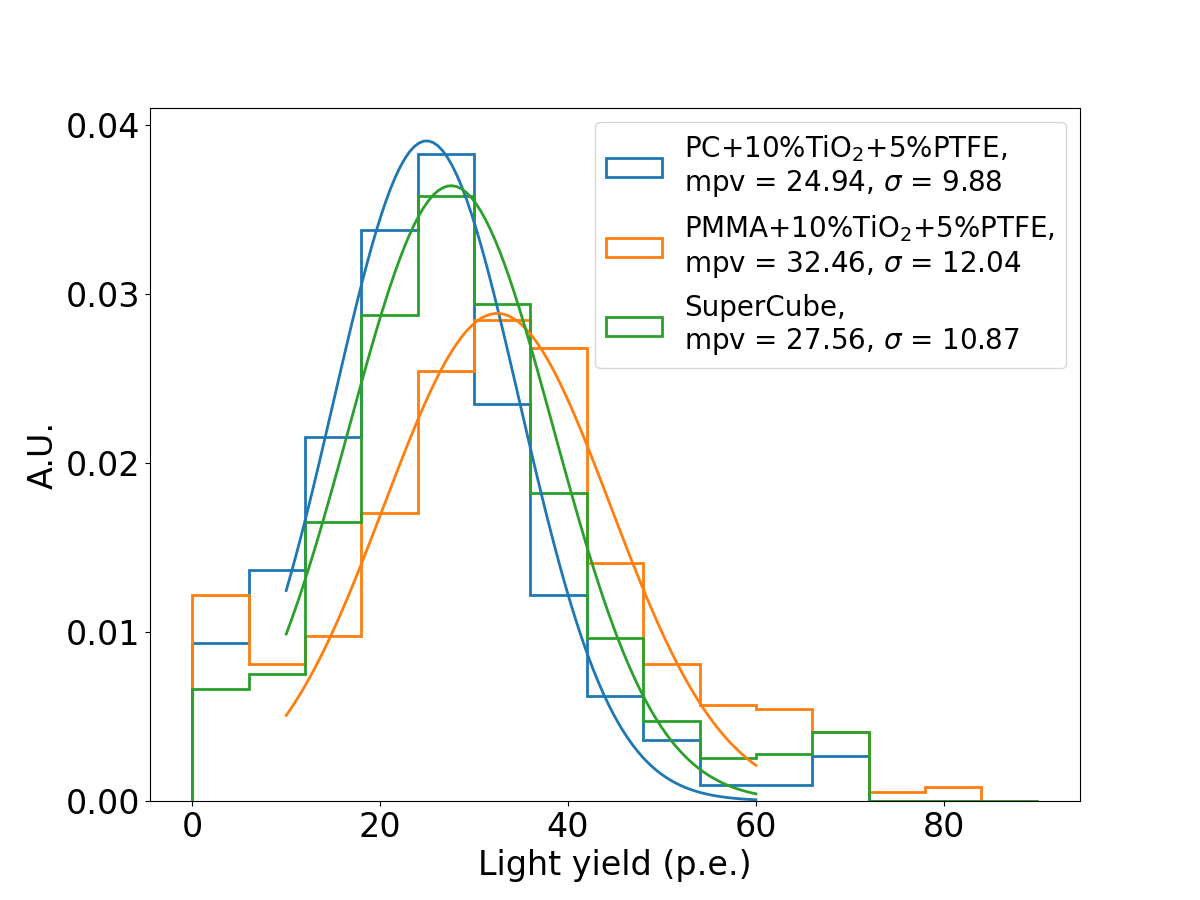}
        \qquad
		\includegraphics[width=0.45\textwidth]{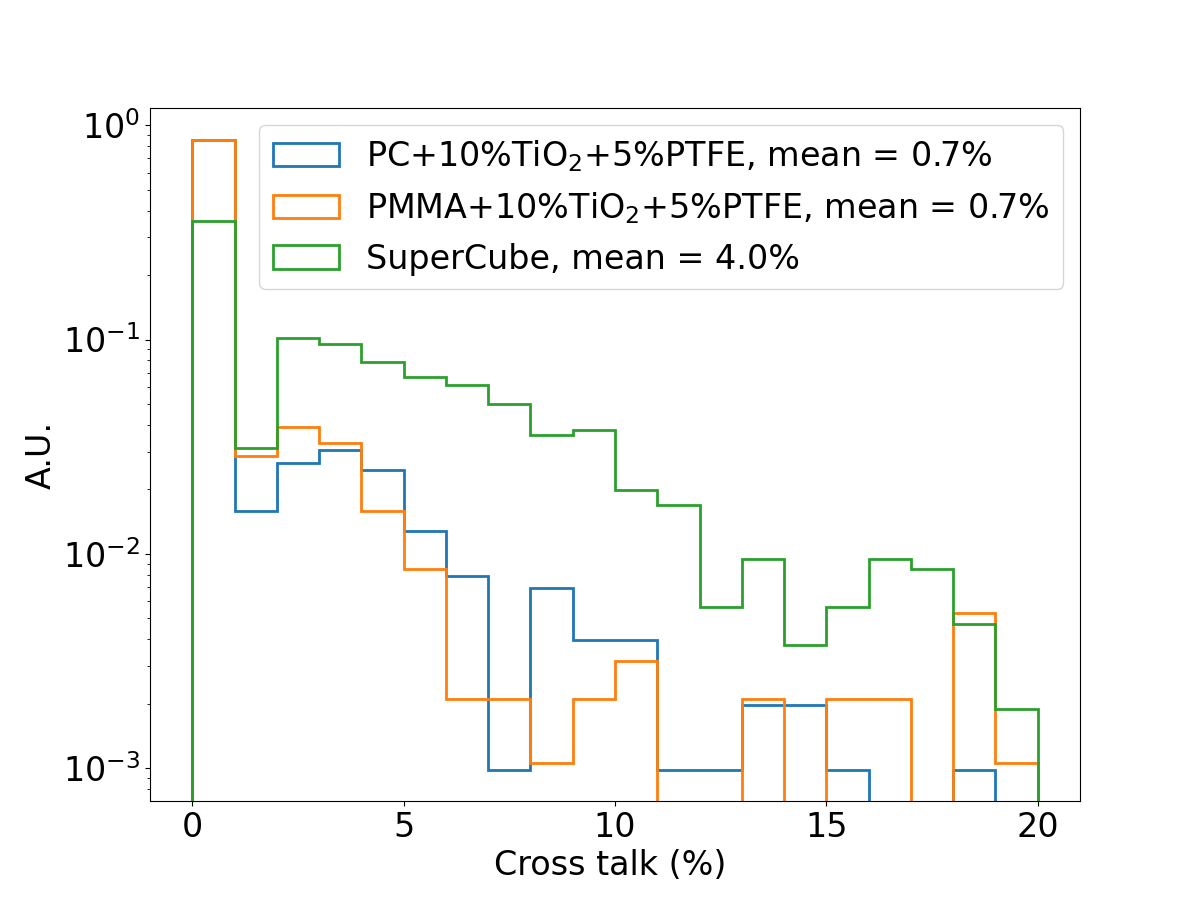}
            \caption{Left: The single channel light yield distributions measured with three different cube layers. Right: The cube-to-cube light crosstalk distributions measured with three different cube layers.
            } 
	\label{fig:measurement_result}
\end{figure}

The single-channel light yield and cube-to-cube light crosstalk were measured to characterise the performance of each prototype. The results for the PC and PMMA filaments mixed with 10\% TiO$_2$ + 5\% PTFE are shown in Fig.~\ref{fig:measurement_result}. 
Each distribution corresponds to the channel-wise detector response to a characterising energy deposition by cosmic rays, 
typically minimum-ionizing particles (MIPs). 
Selecting vertically penetrating muons ensured a path length of around 1~cm in each cube, allowing for a typical energy deposition of approximately 1.8~MeV per cube. 
The light yield was estimated from the most probable value (MPV) obtained with a Gaussian fit.
As expected, the SuperLayer with the PMMA reflector shows the highest light yield MPV of 32.5 p.e./MIP/channel (mean = 32.3 p.e.), compared to the lower 24.9~p.e./MIP/channel (mean = 25.2 p.e.) of the PC-based SuperLayer and the 27.6 p.e./MIP/channel (mean = 29.3) of the SuperCube.

The crosstalk was calculated as the ratio of the light yield measured by two parallel channels: the cube traversed by the selected charged particle and the adjacent cube.
A threshold of 0.7 p.e. was applied to the crosstalk channels, for all readout signals below this threshold (mostly due to the fluctuation of the electronic pedestal) zero crosstalk was assumed. 
The crosstalk distributions are shown in the right panel of Fig.~\ref{fig:measurement_result}. The peaks at zero correspond to the events where the crosstalk is below the threshold. Both the prototypes using PMMA- or PC-based reflectors show low crosstalk, with a mean of 0.7\%. For comparison, the SuperCube, made with the commercial white reflector filament, shows a mean crosstalk of 4\% per face. 

Finally, the SuperLayer made with PMMA + 15\% TiO$_2$ but without PTFE was tested.
The results are summarised in Tab.~\ref{tab:table-reflection}. In the case of the 1~mm thick samples, this filament showed the best optical performance by a small margin, followed by the PMMA + 10\% TiO$_2$ + 5\% PTFE filament.
The measured light yield was 31.2 p.e. and the light crosstalk was 0.7 \%, consistent within the expected experimental uncertainty. 
From the analysis of different runs, the quality of the coupling between the MPPCs and the WLS fibres was found to cause a possible variation of $\sim 1$ p.e..


\section{Conclusions}
\label{sec:conclusions}

In this work, we developed and characterised new custom white reflective filaments for use in 3D printing of finely-segmented plastic scintillator detectors. Two base polymers, polycarbonate (PC) and polymethyl methacrylate (PMMA), were evaluated in combination with reflective additives such as titanium dioxide (TiO$_2$) and polytetrafluoroethylene (PTFE). 

Optical measurements showed that for PC-based filaments, the optimal formulation consisted of 10\% TiO$_2$ + 5\% PTFE, yielding a reflectivity of 85.5\% at 420~nm and a transmittance as low as 0.04\% for a 1.0~mm thick sample. 
For PMMA-based filaments, 15\% TiO$_2$ was sufficient to achieve even higher reflectivity, reaching 92.0\% for a thickness of 0.2~mm. The addition of PTFE to PMMA did not improve the performance, likely due to the inherently lower light absorption and higher transparency of the PMMA matrix. 
In such optically-transparent environments, multi-component formulations may not yield additional benefit, although this outcome may also depend on specific 3D printing conditions and processing parameters.

Complementary measurements were obtained by exposing the 3D printed plastic scintillator prototypes, the new SuperLayers and the old SuperCube, to cosmic rays.
The PMMA-based SuperLayer built with 10\% TiO$_2$ + 5\% PTFE reflector achieved the highest light yield, measuring 32.5 p.e./MIP/channel, outperforming both the PC-based reflector prototype (24.9~p.e./MIP/channel) and the SuperCube reference detector (27.6 p.e./MIP/channel), 
manufactured with the commercial filament that provides the best reflectivity among those that we could investigate.
PMMA with only 15\% TiO$_2$ showed a light yield slightly lower than that of the PMMA-based filament with 10\% TiO$_2$ + 5\% PTFE, showing a slightly different trend compared to the reflectivity measurement, though believed to be within the experimental uncertainties. 
On the other hand, both PMMA- and PC-based SuperLayers exhibited excellent optical isolation, with average cube-to-cube light crosstalk always below 1\%. 

Overall, the results confirm that the newly developed 
PMMA-based reflective filament showed the most-promising performance among the formulations that we investigated in this study.
It improves both the light yield and crosstalk performance of our previous work \cite{supercube-weber,Li:2024txz}, whilst also addressing the requirements for high-resolution calorimetry and particle tracking applications.

\acknowledgments

This work was supported by the joint grant IZURZ2\_224819 of the Swiss National Science Foundation (SNSF) and the National Research Foundation of Ukraine (NRFU).  
This work was also supported by the SNSF grant PCEFP2\_203261.




\bibliographystyle{JHEP}
\bibliography{biblio}

\appendix
\section{Contributions}
A.K., T.S., A.B., B.G., N.K., S.M., M.S. conceptualized, developed and produced the new filament, performed the formal analysis, made the test samples for transmittance and reflectivity tests and performed the measurements. They also provided the equipment at ISMA. 
T.W. 3D printed the SuperLayer prototypes.
B.L. and T.W. built the measurement setup.
B.L. also took the cosmic ray data and performed the data analysis. 
D.S. and U.K. supervised the implementation process and the data analysis. 
A.Ru. provide equipment for measurements at ETHZ.
S.B., E.B., and S.H. provided equipment for the 3D printing of the SuperLayer.
All the authors are part of the 3D-printed DETector (3DET) R\&D collaboration and discussed results and commented on the manuscript.

\end{document}